\documentclass[aps,prd,preprint,a4paper,showpacs,nofootinbib,superscriptaddress]{revtex4-2}
\usepackage{bm}
\usepackage{indentfirst}
\usepackage{amsmath}
\usepackage{graphicx}
\usepackage{float}
\usepackage{amssymb}
\usepackage{subfigure}
\usepackage{amssymb}
\usepackage{hyperref}
\usepackage{epstopdf}
\usepackage[section]{placeins}

\usepackage[utf8]{inputenc}
\hypersetup{
    colorlinks=true,
    linkcolor=red,
    citecolor=blue,
}
\usepackage{color}
\usepackage[T1]{fontenc}
\usepackage{txfonts}
\usepackage{orcidlink}

\usepackage{adjustbox,lipsum}

\newcommand{\Rmnum}[1]{\expandafter\@slowromancap\romannumeral #1@} 
\newcommand{\bq}{\begin{equation}}
\newcommand{\eq}{\end{equation}}
\newcommand{\bqn}{\begin{eqnarray}}
\newcommand{\eqn}{\end{eqnarray}}
\newcommand{\nb}{\nonumber}
\newcommand{\lb}{\label}

\makeatother

\begin{document}

\title{Regge poles, grey body factors, and absorption cross sections for black hole metrics with discontinuity}

\author{Guan-Ru Li}
\affiliation{Faculdade de Engenharia de Guaratinguet\'a, Universidade Estadual Paulista, 12516-410, Guaratinguet\'a, SP, Brazil}

\author{Wei-Liang Qian}
\email{wlqian@usp.br (corresponding author)}
\affiliation{Escola de Engenharia de Lorena, Universidade de S\~ao Paulo, 12602-810, Lorena, SP, Brazil}
\affiliation{Faculdade de Engenharia de Guaratinguet\'a, Universidade Estadual Paulista, 12516-410, Guaratinguet\'a, SP, Brazil}
\affiliation{Center for Gravitation and Cosmology, College of Physical Science and Technology, Yangzhou University, Yangzhou 225009, China}

\author{Qiyuan Pan}
\affiliation{Key Laboratory of Low Dimensional Quantum Structures and Quantum Control of Ministry of Education, Synergetic Innovation Center for Quantum Effects and Applications, and Department of Physics, Hunan Normal University, Changsha, Hunan 410081, China}
\affiliation{Center for Gravitation and Cosmology, College of Physical Science and Technology, Yangzhou University, Yangzhou 225009, China}

\author{Ramin G. Daghigh}
\affiliation{Natural Sciences Department, Metropolitan State University, Saint Paul, Minnesota, 55106, USA}

\author{Jodin C. Morey}
\affiliation{Le Moyne College, Syracuse, NY, 13214-1301, USA}

\author{Rui-Hong Yue}
\affiliation{Center for Gravitation and Cosmology, College of Physical Science and Technology, Yangzhou University, Yangzhou 225009, China}

\begin{abstract}
It was recently proposed by Rosato {\it et al.} and Oshita {\it et al.} that the black hole greybody factors, as stable observables at relatively high frequencies, are more relevant quantities than quasinormal modes in modeling ringdown spectral amplitudes.
The proposed alternative was primarily motivated by the observed spectral instability of quasinormal modes, which was shown to be highly sensitive to ultraviolet metric perturbations. 
It was argued that the overall contributions of spectrally unstable quasinormal modes conspire to produce stable observables through collective interference effects.
Regge poles, which govern the analytic structure of the $S$-matrix in the complex angular momentum plane, also serve as an effective tool for assessing the scattering processes of black holes.
In this regard, the present study investigates the Regge poles and the total absorption cross sections for black hole metrics with discontinuity.
Specifically, these quantities are evaluated at given frequencies and scattering angles, which account for the contributions of the greybody factors at different angular momenta.
To this end, we generalize the matrix method to evaluate the Regge poles in black hole metrics with discontinuity.
To ascertain our approach, the numerical results are compared with those obtained using a modified version of the continued fraction method.
The obtained Regge pole spectrum is then used to calculate the scattering amplitude and cross-section.
We show that the stability of these observables as functions of the frequency can be readily interpreted in terms of the stability of the Regge pole spectrum, particularly the low-lying modes.
Nonetheless, destabilization still occurs at larger frequencies, characterized by the emergence of a bifurcation in the spectrum.
The latter further evolves, leading to more significant deformation in the Regge poles, triggered by ultraviolet metric perturbations moving further away from the black hole.
However, based on the validity of the WKB approximation, it is argued that such an instability in the spectrum is not expected to cause significant observable implications.
\end{abstract}

\date{Sept. 1st, 2025}

\maketitle


\newpage
\section{Introduction}\label{sec1}

Characterizing black hole dynamics through gravitational wave observables has entered a critical phase of quantitative analysis with the advent of interferometric detectors. 
A crucial topic to this endeavor is black hole spectroscopy, which involves modeling of ringdown spectral amplitudes, typically dominated by quasinormal modes (QNMs)~\cite{agr-qnm-review-01, agr-qnm-review-02, agr-qnm-review-03, agr-qnm-review-05, agr-BH-spectroscopy-review-04}. 

As first pointed out by Leaver~\cite{agr-qnm-21, agr-qnm-29}, the QNMs can be attributed to the poles of the frequency-domain Green's function of the underlying master equation that governs the dynamics of the perturbations, whereas the late-time tail is associated with the branch cut on the imaginary frequency axis~\cite{agr-qnm-tail-01}. 
The information inferred from the analytic properties of the Green's function provides a unique avenue for understanding the characteristics of the resulting waveforms.
Recently, the notion of spectral instability has aroused much attention.
As pioneered by Nollert and Price~\cite{agr-qnm-35, agr-qnm-36} and Aguirregabiria and Vishveshwara~\cite{agr-qnm-27, agr-qnm-30}, it was shown that even insignificant perturbations, such as a minor discontinuity~\cite{agr-qnm-instability-11, agr-qnm-lq-03}, can qualitatively deform the higher overtones in the QNM spectrum.
This concept was elaborated and primarily established by Jaramillo {\it et al.}~\cite{agr-qnm-instability-07, agr-qnm-instability-13} by systematically analyzing the impact on the QNM spectrum of randomized and sinusoidal perturbations to the metrics, and demonstrating the prone to instability, particularly against the {\it ultraviolet}, namely, small-scale or high-frequency perturbations.

These findings challenge the conventional assumption that an insignificant modification of the effective potential shall not introduce a sizable deviation in the resulting QNMs. 
Conversely, even in the presence of a rather moderate discontinuity~\cite{agr-qnm-instability-11, agr-qnm-lq-03, agr-qnm-echoes-20}, the asymptotic behavior of the QNM spectrum could be non-perturbatively modified. 
Specifically, high-overtone modes usually lie parallel along the real frequency axis rather than ascending the imaginary frequency axis observed for most black hole metrics~\cite{agr-qnm-continued-fraction-12, agr-qnm-continued-fraction-23}. 
This phenomenon persists regardless of the discontinuity's distance from the horizon or its magnitude.
Specifically, the study of spectral instability and its ramifications has significant observational implications, especially in the context of black hole spectroscopy~\cite{agr-bh-spectroscopy-05, agr-bh-spectroscopy-06, agr-bh-spectroscopy-15, agr-bh-spectroscopy-18, agr-bh-spectroscopy-20, agr-bh-spectroscopy-36}. 
In real-world astrophysical contexts, gravitational radiation sources such as black holes or neutron stars are seldom isolated; they are typically submerged and interacting with surrounding matter. 
The resulting deviations from the ideal symmetric metrics cause the ringdown gravitational waves associated with the QNMs to differ substantially from those predicted for a pristine, isolated, compact object. 

The related topic of spectral instability, echoes, and causality has been explored extensively in recent years by many authors~\cite{agr-qnm-instability-08, agr-qnm-instability-13, agr-qnm-instability-14, agr-qnm-instability-15, agr-qnm-instability-16, agr-qnm-instability-18, agr-qnm-instability-19, agr-qnm-instability-26, agr-qnm-echoes-22, agr-qnm-echoes-29, agr-qnm-echoes-30, agr-qnm-instability-29, agr-qnm-instability-32, agr-qnm-instability-33, agr-qnm-instability-43, agr-qnm-echoes-35}.
Notably, Cheung {\it et al.}~\cite{agr-qnm-instability-15} pointed out that even the fundamental mode can be destabilized under generic perturbations.
Specifically, a small perturbation to the Regge-Wheeler effective potential causes the fundamental mode to spiral outwardly while the deviation's magnitude increases.
Such an observation undermines the understanding that spectral instability might not significantly impact black hole spectroscopy.
This is because it was understood that the fundamental mode is not subject to spectral instability.
The fact that the fundamental mode might not be stable can have substantial observational implications~\cite{agr-qnm-instability-13, agr-qnm-instability-16, agr-qnm-instability-66}.
The physical significance of such instability has been further scrutinized, inclusively regarding an interplay between the distance of the perturbation from the black hole and the size of the physical system, in more recent studies~\cite{agr-qnm-instability-47, agr-qnm-instability-56, agr-qnm-instability-59, agr-qnm-instability-55, agr-qnm-instability-65}.

However, it was proposed recently by Rosato {\it et al.}~\cite{agr-qnm-instability-60} and Oshita {\it et al.}~\cite{agr-qnm-instability-61} that the black hole greybody factors are more robust observables. 
These authors pointed out that greybody factors remain largely stable against small perturbations to the metric until relatively high frequencies, unlike QNMs, making them particularly relevant for interpreting late-time ringdown signals.
It was speculated that the aggregate contributions of unstable QNMs appear to yield stable observables through some collective interference effects. 
Such behavior suggests a subtle decoupling between the impact of individual modes and their collective contribution and, subsequently, warrants deeper investigation.
Specifically, at higher frequencies, more significant deviations in the greybody factors from their unperturbed counterparts are observed, which can be readily understood by the asymptotical values obtained using the WKB approximation~\cite{agr-qnm-instability-61}.
Nonetheless, it was argued that this frequency region lies beyond the relevant frequency bands for the ringdown signals.

From an empirical perspective, regarding the measurements of a scattering process, a pertinent observable is the absorption cross-section, which can also be evaluated at a given frequency.
By definition, it receives contributions from the greybody factors for all different partial waves denoted by their angular momentum $\ell$.
In the context of gravitational wave measurements, it might not be straightforward to distinguish a specific partial wave with a given angular momentum from the others.
Along this train of thought, efforts have been carried out by employing Regge poles~\cite{qft-smatrix-Regge-04}.
Such studies were initiated by Chandrasekhar and Ferrari~\cite{agr-qnm-Regge-01} and Andersson and Thylwe~\cite{agr-qnm-Regge-02} and recently revitalized by D\'ecanini, Esposito-Far\`ese, Folacci {\it et al.}~\cite{agr-qnm-Regge-10,agr-qnm-Regge-11,agr-qnm-Regge-12}.
As singularities in the complex angular momentum plane arising from the analytic continuation of the $S$-matrix, the Regge poles offer a compelling alternative to describe the ringdown waveforms~\cite{agr-qnm-Regge-01, agr-qnm-Regge-02}. 
In particular, an analysis of the absorption cross-section has been carried out by Torres~\cite{agr-qnm-Regge-13} in the context of spectral instability.
Rather distinct from the low-lying QNMs~\cite{agr-qnm-instability-15}, it was observed that the first few Regge poles remain largely unaffected by the metric perturbations.
The latter was shown to be essential in capturing a significant fraction of the resultant cross-section.
In this regard, it is also rather interesting to explore the topic regarding a physically simplified and mathematically attainable scenario, namely, metric perturbation implemented by introducing a discontinuity, which has been shown effective in addressing the instability of both the fundamental~\cite{agr-qnm-instability-55} as well as asymptotic~\cite{agr-qnm-lq-03, agr-qnm-lq-matrix-12} QNMs. 

The present study is motivated by the above considerations.
To this end, we generalize the matrix method~\cite{agr-qnm-lq-matrix-01, agr-qnm-lq-matrix-02} to evaluate the Regge poles in black hole metrics with discontinuities. 
The obtained Regge pole spectrum is used to calculate the scattering amplitude and cross-section. 
Three key results emerge:
\begin{itemize}
\item Observed stability in the scattering amplitude and cross-section at moderate frequencies can be primarily attributed to the stability of the Regge pole spectrum, particularly the first few modes, consistent with the findings of~\cite{agr-qnm-Regge-13, agr-qnm-instability-60, agr-qnm-instability-61}.
\item High-frequency deviation from the original black hole, if any, primarily comes from the first (lowest-lying) Regge pole, in agreement with the WKB approximation.
\item Spectral instability still arises from a more significant deformation of the Regge pole spectrum, triggered by ultraviolet metric perturbations placed further away from the compact object.
\end{itemize}

The remainder of the paper is organized as follows.
In Sec.~\ref{sec2}, we review and generalize the matrix method to evaluate the Regge poles in black hole metrics with discontinuities. 
Subsequently, in Sec.~\ref{sec3}, we evaluate the Regge poles, and the numerical results are ascertained by comparing them with those obtained by a modified continued fraction method.
In Sec.~\ref{sec4}, we aim to explore the resulting scattering amplitude and cross-section and elaborate on the stability of these quantities regarding the spectral instability in the Regge trajectory.
The last section includes further discussions and concluding remarks.

\section{Generalized matrix method for Regge poles}\label{sec2}

In this section, we first give a brief account of the formalism for QNMs and Regge poles.
Subsequently, we generalize the matrix method to evaluate Regge poles, particularly for effective potentials with discontinuity.

The study of black hole perturbation theory often leads to exploring the solution of the radial part of the master equation~\cite{agr-qnm-review-01, agr-qnm-review-02},
\begin{eqnarray}
\frac{\partial^2}{\partial t^2}\phi_{\ell}(t, r_*)+\left(-\frac{\partial^2}{\partial r_*^2}+V_\mathrm{eff}\right)\phi_{\ell}(t, r_*)=0 ,
\label{master_eq_ns}
\end{eqnarray}
where the spatial coordinate $r_*$ is known as the tortoise coordinate, and the effective potential $V_\mathrm{eff}$ is governed by the given spacetime metric.
For instance, the Regge-Wheeler potential $V_\mathrm{RW}$ for the Schwarzschild black hole metric is~\cite{agr-qnm-01}
\bqn
V_\mathrm{eff} = V_\mathrm{RW}=F\left[\frac{\ell(\ell+1)}{r^2}+(1-{\bar{s}}^2)\frac{r_h}{r^3}\right],
\lb{Veff_RW}
\eqn
where ${\bar{s}}$ is the spin and $\ell$ is the multipole number of the angular momentum of the perturbations.
Also, 
\bqn
F=1-r_h/r ,
\lb{f_RW}
\eqn
and $r_h=2M$ is the event horizon radius determined by the black hole's ADM mass, $M$.  
Here, we use geometric units with $G=c=1$.
The tortoise coordinate $r_*\in(-\infty,+\infty)$ is related to the radial coordinate $r\in [0,+\infty)$ by the relation $r_*=\int dr/F(r)$.

In the frequency domain, Eq.~\eqref{master_eq_ns} possesses the form 
\begin{equation}
\frac{d^2\phi_{\ell}(\omega, r_*)}{dr_*^2}+[\omega^2-V_\mathrm{eff}]\phi_{\ell}(\omega, r_*) = 0 . \label{master_frequency_domain}
\end{equation}
It is noted that the problem can be viewed as a one-dimensional scattering process against the effective potential $V_\mathrm{eff}$ with the incident plane wave $\phi_{\ell}=e^{-i\omega r_*}$ coming from spatial infinity $r\to \infty$.
Asymptotically, one denotes the amplitudes of the reflection and transmission waves by $R_\ell=R_\ell(\omega)$ and $T_\ell=T_\ell(\omega)$, and we have
\begin{equation}
\phi_{\ell}^\mathrm{in} \sim
\begin{cases}
   T_\ell e^{-i\omega r_*}, &  r_* \to -\infty, \\
   e^{-i\omega r_*} + R_\ell e^{+i\omega r_*} . &  r_* \to +\infty .
\end{cases}
\label{master_bc_in}
\end{equation}
The counterpart of Eq.~\eqref{master_bc_in}
\begin{equation}
\phi_{\ell}^\mathrm{out} \sim
\begin{cases}
   \widetilde{T}_\ell e^{+i\omega r_*}, &  r_* \to +\infty, \\
   e^{+i\omega r_*} + \widetilde{R}_\ell e^{-i\omega r_*} . &  r_* \to -\infty,
\end{cases}
\label{master_bc_out}
\end{equation}
is another solution that satisfies the master equation, corresponding to the scattering of an outgoing plane wave from the horizon.
The reflection and transmission amplitudes in Eqs.~\eqref{master_bc_in} and~\eqref{master_bc_out} are related by
\bqn
\widetilde{T}_\ell &=& {T}_\ell ,\nb\\
\widetilde{R}_\ell &=& -{R}^*_\ell ,\lb{conjCond}
\eqn
owing to completeness and flux conservation~\cite{book-blackhole-Frolov}.
Until this point, the angular momentum $\ell$ is a non-negative integer, and the frequency $\omega$ is a real number.

The black hole QNMs~\cite{agr-qnm-review-02} are determined by solving the eigenvalue problem defined by Eq.~\eqref{master_frequency_domain} by the following boundary conditions in asymptotically flat spacetimes
\begin{equation}
\phi_{\ell} \sim
\begin{cases}
   e^{-i\omega_{n} r_*}, &  r_* \to -\infty, \\
   e^{+i\omega_{n} r_*}, &  r_* \to +\infty,
\end{cases}
\label{master_bc0}
\end{equation}
for which, in order to satisfy the above boundary conditions, the frequency will take discrete complex values $\omega_n$, known as the quasinormal frequencies, where the subscript $n$ is referred to as the overtone number.
By comparing Eq.~\eqref{master_bc0} against Eq.~\eqref{master_bc_in} or~\eqref{master_bc_out}, it is apparent that the former is attained when the reflection coefficient becomes divergent~\cite{agr-qnm-Poschl-Teller-02}.
Moreover, the Wronskian between the wave functions Eqs.~\eqref{master_bc_in} and~\eqref{master_bc_out} vanishes,
\bqn
W\left(\phi_{\ell}^\mathrm{in}, \phi_{\ell}^\mathrm{out}\right) = \phi_{\ell}^\mathrm{in}{\phi'}_{\ell}^\mathrm{out}-\phi_{\ell}^\mathrm{out}{\phi'}_{\ell}^\mathrm{in} = 0
\eqn
as the two solutions become linearly dependent, where the prime is the derivative with respect to the spatial coordinate.

The Regge poles~\cite{agr-qnm-Regge-01, agr-qnm-Regge-02} are defined as the poles of the reflection amplitudes $R_\ell$ in the analytically continued angular momentum space, evaluated at a given (real-valued) frequency while (formally) adopting the same boundary condition Eq.~\eqref{master_bc0}.
Its resemblance to the QNMs is readily recognized as the boundary condition implies divergence in the scattering matrix as the reflection wave overwhelms the incident one.
Subsequently, it leads to a vanishing Wronskian.
Because the asymptotic waveforms given by Eqs.~\eqref{master_bc_in} and~\eqref{master_bc_out} are linearly dependent, a pole in $R_\ell$ thus implies a pole in $T_\ell$ and vice versa\footnote{It is worth noting that the flux conservation condition $|R|^2+|T|^2=1$ is no longer valid for complex angular momentum, which also invalidates Eq.~\eqref{conjCond}.}.
Similar to the QNMs, it is usually assumed in the literature that these poles are simple ones, and one denotes the residue for the $n$th pole as
\bqn
r_n \equiv \mathrm{Res} [e^{i\pi(\ell+1)}R_\ell(\omega)]_{\ell=\ell_n} .
\eqn
For the remainder of this paper, we will exclusively use the subscript $n$ to refer to the index of Regge poles.
The main advantage of making use of the Regge poles is due to the relationship between the scattering amplitude and reflection coefficient~\cite{book-quantum-mechanics-Sakurai} and the Watson-Sommerfeld transform~\cite{book-methods-mathematical-physics-10}, which provides a means for converting potentially slowly converging series into contour integrals using the Cauchy residue theorem.
Specifically, the greybody factor $\Gamma_\ell$~\cite{agr-bh-superradiance-01, agr-bh-superradiance-02}, scattering amplitude $f$, the differential $d\sigma_\mathrm{abs}/d\Omega$ and total cross-section $\sigma_\mathrm{abs}$~\cite{agr-qnm-Regge-02, agr-qnm-11} satisfy the following forms
\bqn
f(\omega, \theta) = f^\mathrm{RP}(\omega, \theta) + f^\mathrm{BG}(\omega, \theta) ,
\eqn
where
\bqn
f^\mathrm{RP}(\omega, \theta) = -\frac{i\pi}{\omega}\sum_{n=0}^{\infty} \frac{\lambda_n(\omega) r_n(\omega)}{\cos\pi\lambda_n(\omega)}P_{\ell_n(\omega)}(-\cos\theta) .\lb{fRP}
\eqn
with
\bqn
\lambda \equiv \ell +\frac12 .
\eqn
Here, $\lambda_n$ and $\ell_n$ are evaluated at $n$th Regge pole.
The order of the Legendre polynomials $P_\ell$, inherited from the partial wave formula, is also governed by the Regge poles as a result of the contour integration.
The differential cross-section is given by the scattering amplitude
\bqn
\frac{d\sigma_\mathrm{abs}}{d\Omega} = \left|f(\omega, \theta)\right|^2 ,
\eqn
and
\bqn
\sigma_\mathrm{abs}(\omega) \equiv \int \frac{d\sigma_\mathrm{abs}}{d\Omega} d\Omega = \sigma_\mathrm{abs}^\mathrm{RP}(\omega) + \sigma_\mathrm{abs}^\mathrm{BG}(\omega) ,
\eqn
where
\bqn
\sigma_\mathrm{abs}^\mathrm{RP}(\omega) = -\frac{4\pi^2}{\omega^2}\mathrm{Re}\left[\sum_{n=0}^{\infty}\frac{\lambda_n(\omega)\gamma_n(\omega)e^{i\pi\ell_n(\omega)}}{\sin\pi\ell_n(\omega)}\right], \lb{sigmaRP}
\eqn
where $\gamma_n$ is the residue of the greybody factor
\bqn
\gamma_n(\omega) \equiv \mathrm{Res}\left[\Gamma_{\lambda-1/2}(\omega)\right]_{\lambda=\lambda_n}
\eqn
whereas the greybody factor is defined as
\bqn
\Gamma_\ell(\omega) \equiv \Gamma_{\lambda-1/2}(\omega) =\left|T_\ell(\omega)\right|^2  ,
\eqn
The background contributions $f^\mathrm{BG}$ and $\sigma_\mathrm{abs}^\mathrm{BG}$ come from the high-energy limit of the scattering process and can be estimated using means such as the WKB approximation (also see further discussions in Sec.~\ref{sec4}).

By observing Eq.~\eqref{fRP} and~\eqref{sigmaRP}, it is observed that both the scattering amplitude and the total cross-section are constituted by contributions from partial waves with different angular momenta.
In other words, these quantities can be readily evaluated once the Regge poles and the corresponding residues are determined.
In what follows, we generalize the matrix method, which was initially developed to evaluate QNMs to compute the Regge poles.

The {\it matrix method}~\cite{agr-qnm-lq-matrix-01} is inspired by one of the most accurate numerical approaches to date for evaluating the QNMs, the continued fraction method~\cite{agr-qnm-continued-fraction-01, agr-qnm-continued-fraction-03, agr-qnm-continued-fraction-04}.
While appropriately taking into account the asymptotic wave function, such an approach expands the wave function at a given coordinate and reiterates the master equation~\eqref{master_eq_ns} as an iterative relation between the expansion coefficients, which can be equivalently expressed in a mostly diagonal matrix form.
The QNMs problem is subsequently solved by using Hill's determinant.
Instead of a given position, the matrix method discretizes the entire spatial domain and carries out the expansions of the wave function on the entire grid~\cite{agr-qnm-lq-matrix-01}.
The master equation is, therefore, formulated into a mostly dense matrix equation, and the QNM problem is reiterated as an algebraic nonlinear equation for the complex frequencies.
Besides the metrics with spherical symmetry~\cite{agr-qnm-lq-matrix-02}, the matrix method can be applied to black hole spacetimes with axial symmetry~\cite{agr-qnm-lq-matrix-03} and a system composed of coupled degrees of freedom~\cite{agr-qnm-lq-matrix-07}.
The approach is shown to be effective in dealing with different boundary conditions~\cite{agr-qnm-lq-matrix-04} and dynamic black holes~\cite{agr-qnm-lq-matrix-05}.
More recently, the original method was generalized to handle effective potentials containing discontinuity~\cite{agr-qnm-lq-matrix-06} and pushed to higher orders~\cite{agr-qnm-lq-matrix-08}, inclusively implemented on Chebyshev grids~\cite{agr-qnm-lq-matrix-11}.
The latter is related to the notorious drawback of the equispaced interpolation points.
Specifically, polynomial interpolation based on a uniform grid is often liable to Runge's phenomenon, characterized by significant oscillations at the edges of the relevant interval, closely related to an increasing Lebesgue constant~\cite{book-approximation-theory-Rivlin, agr-qnm-lq-matrix-10}.
In other words, even though the uniform convergence over the interval in question is provided, an interpolation of a higher degree does not necessarily guarantee an improved accuracy, similar to the Gibbs phenomenon in Fourier series approximations.
On top of the above recipe, another crucial ingredient is the hyperboloidal coordinates~\cite{agr-qnm-hyperboloidal-03}.
In the place of spatial infinity adopted by most approaches as the boundary, the hyperboloidal coordinates guarantee that the boundary condition of the problem is evaluated at null infinity.
This generalization was implemented recently~\cite{agr-qnm-lq-matrix-12} to discuss the spectral instability in the modified P\"oschl-Teller effective potential.

In what follows, we discuss how the matrix method can be revised to evaluate the Regge poles.
Similar to the original version of the matrix method, one rewrites the master equation Eq.~\eqref{master_eq_ns} from $(t, r_*)$ into the compactified hyperboloidal coordinates $(\tau, x)$, where $x\in [-1,1]$~\cite{agr-qnm-hyperboloidal-03}.
Following Ref.~\cite{agr-qnm-instability-07}, one transforms the coordinates into dimensionless quantities $(\overline{t}, \overline{x})$ by introducing a length scale $\xi$. 
Specifically,
\bqn
\overline{t}=\frac{t}{\xi},~~~~\overline{x}=\frac{r_{*}}{\xi},~~~~{\hat{V}_\mathrm{eff}=\xi^2 V_\mathrm{eff}},
\lb{dimensionless_quantities}
\eqn
where the choice of $\xi$ is rather arbitrary, primarily aimed to simplify the resultant master equation. 
Subsequently, the compactified hyperboloidal coordinates $(\tau,x)$ are defined by
\bqn
\overline{t}=\tau-H(\overline{x}),~~~~\overline{x}=G(x).
\lb{compactified_hyperboloidal_approach}
\eqn
where the function $G(x)$ introduces a spatial compactification while the height function $H(x)$ is defined to guarantee that the boundary at $x=\pm 1$ for a given $\tau$ is a null infinity.
Specifically, 
\bqn
\partial_xH &\le& 1 ,\nb\\
\lim\limits_{x\to \pm 1} \partial_xH &=& \pm 1,
\lb{H_relation}
\eqn
where
$\partial_xH \equiv \frac{\partial H\left(G(x)\right)}{\partial x}$.
By plugging the separation of the variables
\bqn
\phi_{\ell}(\tau,x)=e^{-i\omega \xi \tau}\varphi_{\ell}(x),
\lb{wave_function}
\eqn
into the master equation, one finds the counterpart of Eq.~\eqref{master_frequency_domain}
\bqn
\left[a(\omega,x)+b(\omega,x)\partial_{x}+c(\omega,x)\partial_{xx}\right]\varphi_{\ell}=0,
\lb{simplified_master_equation}
\eqn
where
\bqn
a(\omega,x) &=& \xi\omega\left(\frac{i(\partial_{x}H)(\partial_{xx}G)}{(\partial_{x}G)^3}+\frac{\xi\omega(\partial_{x}H)^2-i(\partial_{xx}H)}{(\partial_{x}G)^2}-\xi\omega\right)+\hat{V}_\mathrm{eff},\nb\\
b(\omega,x) &=& \frac{(\partial_{xx}G)-2i\xi\omega(\partial_{x}G)(\partial_{x}H)}{(\partial_{x}G)^3},\nb\\
c(\omega,x) &=& -\frac{1}{(\partial_{x}G)^2},
\eqn
which is an ordinary second-order differential equation in $x$ defined on the interval $[-1,1]$.
It is readily shown that the frequency $\omega$ in Eq.~\eqref{simplified_master_equation} is identical to that defined in Eq.~\eqref{master_frequency_domain}.
Now one discretizes Eq.~\eqref{simplified_master_equation} on a Chebyshev grid~\cite{book-approximation-theory-Rivlin} of $N$ nodes to find a matrix equation~\cite{agr-qnm-lq-matrix-11}
\bqn
\mathcal{M}\ \Vec{\varphi_{\ell}}=0,
\lb{shorthand_master_equation}
\eqn
where the coefficient $\mathcal{M(\omega)}$ is an $N\times N$ matrix whose elements are functions of the frequency and $\vec{\varphi_{\ell}}$ is a $1\times N$ column matrix associated with the wave function,
\bqn
\vec{\varphi_{\ell}}=\left({\varphi_\ell}_0,{\varphi_\ell}_1,\cdots,{\varphi_\ell}_N\right)^{\rm T} ,
\lb{origin_psi}
\eqn
where ${\varphi_\ell}_i\equiv{\varphi_\ell}(x_i)$ are evaluated on the Chebyshev grid
\bqn
x_i={\cos}\left(\frac{i}{N}\pi\right),
\lb{Chebyshev_type_grids}
\eqn
where $i=0,\cdots,N$.

Subsequently, the Regge poles are attained at vanishing determinant
\bqn
\rm det(\mathcal{M})=0.
\lb{origin_solve}
\eqn
We note that Eq.~\eqref{origin_solve} is precisely the same equation proposed in the original matrix method~\cite{agr-qnm-lq-matrix-01}, except that we are now solving for unknown complex angular momenta $\ell_n$ in place of frequency.
This was facilitated by the fact that the form of the boundary condition Eq.~\eqref{master_bc0} remains unchanged.
The elements of the coefficient matrix $\mathcal{M}$ are polynomials in $\ell$ of finite order.
Specifically, this allows us to rewrite the matrix equation in the form
\bqn
\mathcal{M}\ \vec{{\varphi_\ell}}=\left[M_0+M_1\ell+M_2\ell^2\right]\vec{{\varphi_\ell}}=0 ,\lb{mJason}
\eqn
where all polynomials are at most quadratic in $\ell$ for the Regge-Wheeler effective potential Eq.~\eqref{Veff_RW}.
Although Eq.~\eqref{origin_solve} can be solved using the standard numerical methods, it is more convenient to transform it into a linear form~\cite{agr-qnm-59} in $\ell$.
This procedure leads to a $3N\times 3N$ matrix equation for a generalized eigenvalue problem, which is readily solved using \textit{Mathematica}'s subroutines {\it Eigenvalue} or {\it Eigensystem}.

Now, we proceed to generalize the above algorithm for the effective potential that contains discontinuity~\cite{agr-qnm-lq-matrix-06}.
Let us consider that the effective potential is discontinuous at one grid point $x = x_c$.
Such a discontinuity should be taken into account by considering Israel's junction condition~\cite{agr-collapse-thin-shell-03, book-general-relativity-Poisson}.
As the Taylor expansion becomes invalid in the entire domain, the wave functions are restricted to their respective sides of the discontinuity. 
They are related by the condition~\cite{agr-qnm-34}
\bqn
\lb{Israel1}
\lim_{\epsilon\to 0^+}\left[{\varphi(x_c+\epsilon)}-{\varphi(x_c-\epsilon)}\right]=0 \ ,
\eqn
and
\bqn
\lb{Israel}
\lim_{\epsilon\to 0^+}\left[\frac{\varphi'(x_c+\epsilon)}{\varphi(x_c+\epsilon)}-\frac{\varphi'(x_c-\epsilon)}{\varphi(x_c-\epsilon)}\right]=\kappa \ ,
\eqn
at the point of discontinuity, where, for the Schr\"odinger-type master equation Eq.~\eqref{master_frequency_domain},
\bqn
\lb{kappa}
\kappa = \lim_{\epsilon\to 0^+}\int_{x_c-\epsilon}^{x_c+\epsilon} \hat{V}_{\mathrm{eff}}(x)dx \ .
\eqn
If one considers a moderate finite jump, the above relation simplifies to the condition of vanishing Wronskian
\bqn
\lb{WronskianDis}
W_c \equiv \varphi(x_c+\epsilon)\varphi'(x_c-\epsilon) - \varphi(x_c-\epsilon)\varphi'(x_c+\epsilon) = 0~.
\eqn

Although the above condition is bilinear, it can be adequately taken into account by revising the matrix ${\cal M}$ in Eq.~\eqref{origin_solve} by employing the fact that the wave function is continuous at $x=x_c$.
In practice, the matrix ${\cal M}$ is {\it almost} block diagonalized since the Taylor expansions are mainly restrictive within smaller domains.
The relation Eq.~\eqref{Israel} or~\eqref{WronskianDis} can be implemented on the line between two successive blocks by replacing the information on the wave functions' first-order derivatives with the discrete values of the entire wave function on the grid.
Specifically, if the above boundary corresponds to the $i$th grid, then the $i$th line and column of the matrix ${\cal M}$ are occupied by both blocks.
The matrix must be modified so that the $i$th line must correspond to the junction condition between waveforms on the boundary.
For the specific form of Eq.~\eqref{WronskianDis}, the $i$th column is filled with elements of both blocks because the values of the entire waveform $x\in (0,1)$ including both regions, enter the interpolation schemes for those of the $i$th grid.
We note that the matrix ${\cal M}$ and its modified version cannot be entirely block diagonalized; otherwise, the resulting spectrum would be constituted by a simple summation of those pertaining to individual blocks.

\section{Regge poles in modified Regge-Wheeler potentials}\label{sec3}

This section presents the numerical results on the Regge poles in modified Regge-Wheeler potentials for different scenarios.
We first introduce a truncated Regge-Wheeler potential of the following form
\bqn
V^{(1)}_\mathrm{eff} 
= \begin{cases}
   V_\mathrm{RW}(r), &  r \le r_c, \\
   0, &  r > r_c,
\end{cases}
\lb{Veff_MRW1}
\eqn
where $r_c$ is a point of discontinuity, and the Regge-Wheeler potential is given by Eq.~\eqref{Veff_RW}.

Alternatively, the second scenario considers a modified Regge-Wheeler potential where the truncation is inward
\bqn
V^{(2)}_\mathrm{eff} 
= \begin{cases}
   0, &  r \le r_c, \\
   V_\mathrm{RW}(r), &  r > r_c,
\end{cases}
\lb{Veff_MRW2}
\eqn
where $r_c$ is the point of discontinuity.

Lastly, we consider a modified Regge-Wheeler potential with a minor step at the point of discontinuity
\bqn
V^{(3)}_\mathrm{eff} 
= \begin{cases}
   V_\mathrm{RW}(M, r), &  r \le r_c, \\
   V_\mathrm{RW}(M+\Delta M, r), &  r > r_c .
\end{cases}
\lb{Veff_MRW3}
\eqn
Here, for simplicity, we have started from the effective potential, and the corresponding metrics associated with Eqs.~\eqref{Veff_MRW1}-\eqref{Veff_MRW3} should be derived by solving for the unknown function in some metric ansatz.
However, due to the simple forms, one can make use of Birkhoff's theorem for spherically symmetric cases and conclude that Eq.~\eqref{Veff_MRW3} furnishes a simplified scenario that is physically pertinent, corresponding to an infinitely thin mass shell $\Delta M$ wrapping around the black hole.
For the present case, we understand that when the thickness of the mass shell becomes finite, it does not lead to a significant modification of the effective potential that possibly triggers the spectral instability in QNMs.
Nonetheless, the physical significance of perturbations to the metric and effective potential is a crucial topic; a small modification in the effective potential does not necessarily imply that the corresponding metric perturbation is insignificant.
These concerns have been elaborated recently in~\cite{agr-qnm-instability-47}.

\begin{table}[htbp] 
\caption{\label{Tab.1} The obtained low-lying Regge poles $\lambda_{n}(\omega)$ for the modified Regge-Wheeler potential Eq.~\eqref{Veff_MRW1} with a truncation at $x=x_c$ toward the spatial infinity.
It is noted that, for the compactified hyperbolic coordinate used in this study, the horizon and spatial infinity correspond to $x=0$ and $x=1$, respectively.
The calculations are carried out for $M=1/2$ and axial gravitational perturbations $\bar{s}=2$.
The results are obtained for different frequencies $\omega$ and truncation point $x_c$ by using the improved matrix method (IMM) and the continued fraction method (CFM).} 
\adjustbox{max width=\textwidth}{%
\begin{tabular}{cccccccc}
\hline
\hline
      $$&$\omega$&$x_c$ &$1/5$&$1/3$&$1/2$&$2/3$&$4/5$  \\
\hline
$\mathrm{IMM}$&$0.1$&$\mathrm{Re}(\lambda_{0})~$&$1.774253~$&$1.693898~$&$1.617204~$&$1.557917~$&$1.528656~$  \\
$$&$$&$\mathrm{Im}(\lambda_{0})~$&$0.285414~$&$0.182655~$&$0.132764~$&$0.110739~$&$0.102096~$  \\
$$&$$&$\mathrm{Re}(\lambda_{1})~$&$0.451043~$&$0.380904~$&$0.370980~$&$0.440004~$&$0.575100~$  \\
$$&$$&$\mathrm{Im}(\lambda_{1})~$&$3.598320~$&$2.390483~$&$1.503480~$&$0.874645~$&$0.526054~$  \\
\hline
$$&$2.0$&$\mathrm{Re}(\lambda_{0})~$&$6.716861~$&$5.671341~$&$5.256377~$&$5.477237~$&$5.788201~$  \\
$$&$$&$\mathrm{Im}(\lambda_{0})~$&$1.858302~$&$1.204032~$&$0.661783~$&$0.323130~$&$0.594422~$  \\
$$&$$&$\mathrm{Re}(\lambda_{1})~$&$6.928930~$&$5.618690~$&$4.990127~$&$5.071166~$&$5.334647~$  \\
$$&$$&$\mathrm{Im}(\lambda_{1})~$&$4.735250~$&$3.198286~$&$2.044238~$&$1.064024~$&$0.595809~$  \\
\hline
$\mathrm{CFM}$&$0.1$&$\mathrm{Re}(\lambda_{0})~$&$1.774253~$&$1.693899~$&$1.617216~$&$1.558007~$&$1.528872~$  \\
$$&$$&$\mathrm{Im}(\lambda_{0})~$&$0.285414~$&$0.182655~$&$0.132764~$&$0.110729~$&$0.102096~$  \\
$$&$$&$\mathrm{Re}(\lambda_{1})~$&$0.451043~$&$0.380904~$&$0.370986~$&$0.440233~$&$0.575376~$  \\
$$&$$&$\mathrm{Im}(\lambda_{1})~$&$3.598320~$&$2.390481~$&$1.503475~$&$0.875028~$&$0.528817~$  \\
\hline
$$&$2.0$&$\mathrm{Re}(\lambda_{0})~$&$6.716861~$&$5.671342~$&$5.256464~$&$5.476614$&$5.786688~$  \\
$$&$$&$\mathrm{Im}(\lambda_{0})~$&$1.858301~$&$1.204025~$&$0.661678~$&$0.324511~$&$0.593437~$  \\
$$&$$&$\mathrm{Re}(\lambda_{1})~$&$6.928930~$&$5.618684~$&$4.990179~$&$5.071176~$&$5.334340~$  \\
$$&$$&$\mathrm{Im}(\lambda_{1})~$&$4.735252~$&$3.198287~$&$2.044712~$&$1.066662~$&$0.596942~$  \\
\hline
\hline
\end{tabular}}
\end{table}

\begin{table}[htbp] 
\caption{\label{Tab.2} 
Similar to Tab.~\ref{Tab.1}, but for the low-lying Regge poles $\lambda_{n}(\omega)$ for the truncated Regge-Wheeler potential Eq.~\eqref{Veff_MRW2} with a truncation toward the horizon.
The calculations are carried out for $M=1/2$ and axial gravitational perturbations $\bar{s}=2$.} 
\adjustbox{max width=\textwidth}{%
\begin{tabular}{cccccccc}
\hline
\hline
      $$&$\omega$&$x_c$ &$1/5$&$1/3$&$1/2$&$2/3$&$4/5$  \\
\hline
$\mathrm{IMM}$&$0.1$&$\mathrm{Re}(\lambda_{0})~$&$1.360288~$&$1.268653~$&$1.174931~$&$1.146382~$&$1.301967~$  \\
$$&$$&$\mathrm{Im}(\lambda_{0})~$&$0.094145~$&$0.116269~$&$0.176222~$&$0.307970~$&$0.521280~$  \\
$$&$$&$\mathrm{Re}(\lambda_{1})~$&$0.697955~$&$0.766626~$&$0.903650~$&$1.141025~$&$1.542527~$  \\
$$&$$&$\mathrm{Im}(\lambda_{1})~$&$0.764168~$&$0.961770~$&$1.202828~$&$1.487517~$&$1.830079~$  \\
\hline
$$&$2.0$&$\mathrm{Re}(\lambda_{0})~$&$5.340572~$&$5.660748~$&$6.582699~$&$8.656515~$&$12.905628~$  \\
$$&$$&$\mathrm{Im}(\lambda_{0})~$&$0.664842~$&$0.972551~$&$1.344722~$&$1.800766~$&$2.388594~$  \\
$$&$$&$\mathrm{Re}(\lambda_{1})~$&$5.359182~$&$5.908920~$&$7.120976~$&$9.518427~$&$14.122522~$  \\
$$&$$&$\mathrm{Im}(\lambda_{1})~$&$1.924792~$&$2.415844~$&$3.041853~$&$3.850981~$&$4.904038~$  \\
\hline
$\mathrm{CFM}$&$0.1$&$\mathrm{Re}(\lambda_{0})~$&$1.360256~$&$1.268666~$&$1.174929~$&$1.146381~$&$1.301967~$  \\
$$&$$&$\mathrm{Im}(\lambda_{0})~$&$0.094148~$&$0.116276i$&$0.176223~$&$0.307971~$&$0.521280~$  \\
$$&$$&$\mathrm{Re}(\lambda_{1})~$&$0.698675~$&$0.766448~$&$0.903694~$&$1.141281~$&$1.542476~$  \\
$$&$$&$\mathrm{Im}(\lambda_{1})~$&$0.763163~$&$0.961685~$&$1.202493~$&$1.487527~$&$1.830102~$  \\
\hline
$$&$2.0$&$\mathrm{Re}(\lambda_{0})~$&$5.340564~$&$5.660748~$&$6.582699~$&$8.656515~$&$12.905628~$  \\
$$&$$&$\mathrm{Im}(\lambda_{0})~$&$0.664837~$&$0.972551~$&$1.344722~$&$1.800766~$&$2.388594~$  \\
$$&$$&$\mathrm{Re}(\lambda_{1})~$&$5.359185~$&$5.908902~$&$7.120967~$&$9.518454~$&$14.122522~$  \\
$$&$$&$\mathrm{Im}(\lambda_{1})~$&$1.924777~$&$2.415850~$&$3.041877~$&$3.850958~$&$4.904039~$  \\
\hline
\hline
\end{tabular}}
\end{table}

\begin{table}[htbp] 
\caption{\label{Tab.3} 
Similar to Tab.~\ref{Tab.1}, but for the low-lying Regge poles $\lambda_{n}(\omega)$ for the modified Regge-Wheeler potential Eq.~\eqref{Veff_MRW3} with a thin mass shell.
The calculations are carried out for $M=1/2$, $\Delta M=0.0001$, and axial gravitational perturbations $\bar{s}=2$.} 
\adjustbox{max width=\textwidth}{%
\begin{tabular}{cccccccc}
\hline
\hline
      $$&$\omega$&$x_c$ &$1/5$&$1/3$&$1/2$&$2/3$&$4/5$  \\
\hline
$\mathrm{IMM}$&$0.1$&$\mathrm{Re}(\lambda_{0})~$&$1.518548~$&$1.518545~$&$1.518544~$&$1.518544~$&$1.518544~$  \\
$$&$$&$\mathrm{Im}(\lambda_{0})~$&$0.096582~$&$0.096577~$&$0.096576~$&$0.096578~$&$0.096579~$  \\
$$&$$&$\mathrm{Re}(\lambda_{1})~$ &$0.749823~$&$0.749822~$&$0.749820~$&$0.749807~$&$0.749784~$  \\
$$&$$&$\mathrm{Im}(\lambda_{1})~$&$0.393364~$&$0.393364~$&$0.393363~$&$0.393359~$&$0.393350~$  \\
\hline
$$&$2.0$&$\mathrm{Re}(\lambda_{0})~$&$5.436726~$&$5.436357~$&$5.435848~$&$5.435868~$&$5.435808~$  \\
$$&$$&$\mathrm{Im}(\lambda_{0})~$&$0.471288~$&$0.471466~$&$0.471708~$&$0.471361~$&$0.471499~$  \\
$$&$$&$\mathrm{Re}(\lambda_{1})~$&$5.464037~$&$5.463970~$&$5.464071~$&$5.462719~$&$5.464572~$  \\
$$&$$&$\mathrm{Im}(\lambda_{1})~$&$1.414218~$&$1.414427~$&$1.414758~$&$1.413360~$&$1.414582~$  \\
\hline
$\mathrm{CFM}$&$0.1$&$\mathrm{Re}(\lambda_{0})~$&$1.518550~$&$1.518547~$&$1.518545$&$1.518544~$&$1.518544~$  \\
$$&$$&$\mathrm{Im}(\lambda_{0})~$&$0.096582~$&$0.096577~$&$0.096576~$&$0.096578~$&$0.096579~$  \\
$$&$$&$\mathrm{Re}(\lambda_{1})~$&$0.749806~$&$0.749819~$&$0.749822~$&$0.749806~$&$0.749784~$  \\
$$&$$&$\mathrm{Im}(\lambda_{1})~$&$0.393369~$&$0.393359~$&$0.393364~$&$0.393359~$&$0.393350~$  \\
\hline
$$&$2.0$&$\mathrm{Re}(\lambda_{0})~$&$5.436726~$&$5.436357~$&$5.435848~$&$5.435868~$&$5.435809~$  \\
$$&$$&$\mathrm{Im}(\lambda_{0})~$&$0.471288~$&$0.471466~$&$0.471708~$&$0.471361~$&$0.471501~$  \\
$$&$$&$\mathrm{Re}(\lambda_{1})~$&$5.464037~$&$5.463970~$&$5.464071~$&$5.462719~$&$5.464622~$  \\
$$&$$&$\mathrm{Im}(\lambda_{1})~$&$1.414218~$&$1.414427~$&$1.414758~$&$1.413360~$&$1.414584~$  \\
\hline
\hline
\end{tabular}}
\end{table}

\begin{figure}[htp]
\centering
\includegraphics[scale=0.5]{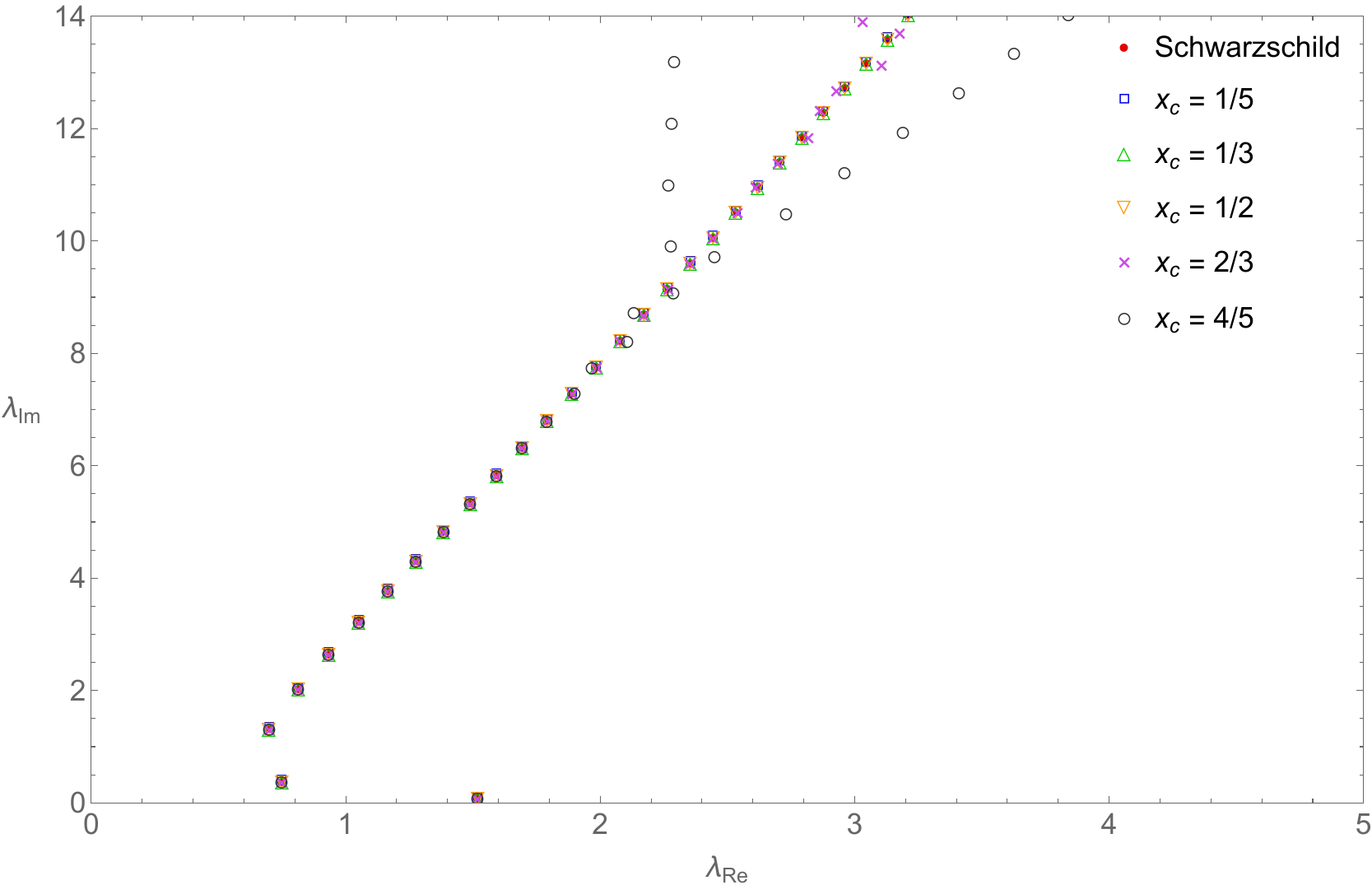}
\caption{The resulting Regge pole spectra of the modified Regge-Wheeler potential Eq.~\eqref{Veff_MRW3} with a minor step placed at different points: $x_c= 1/5, 1/3$, $1/2$, $2/3$, and $4/5$, evaluated for a given frequency $\omega = 0.1$.
The results are compared to the unperturbed Schwarzschild metric.}
\label{fig01_RP_RWout}
\end{figure}

\begin{figure}[htp]
\centering
\includegraphics[scale=0.5]{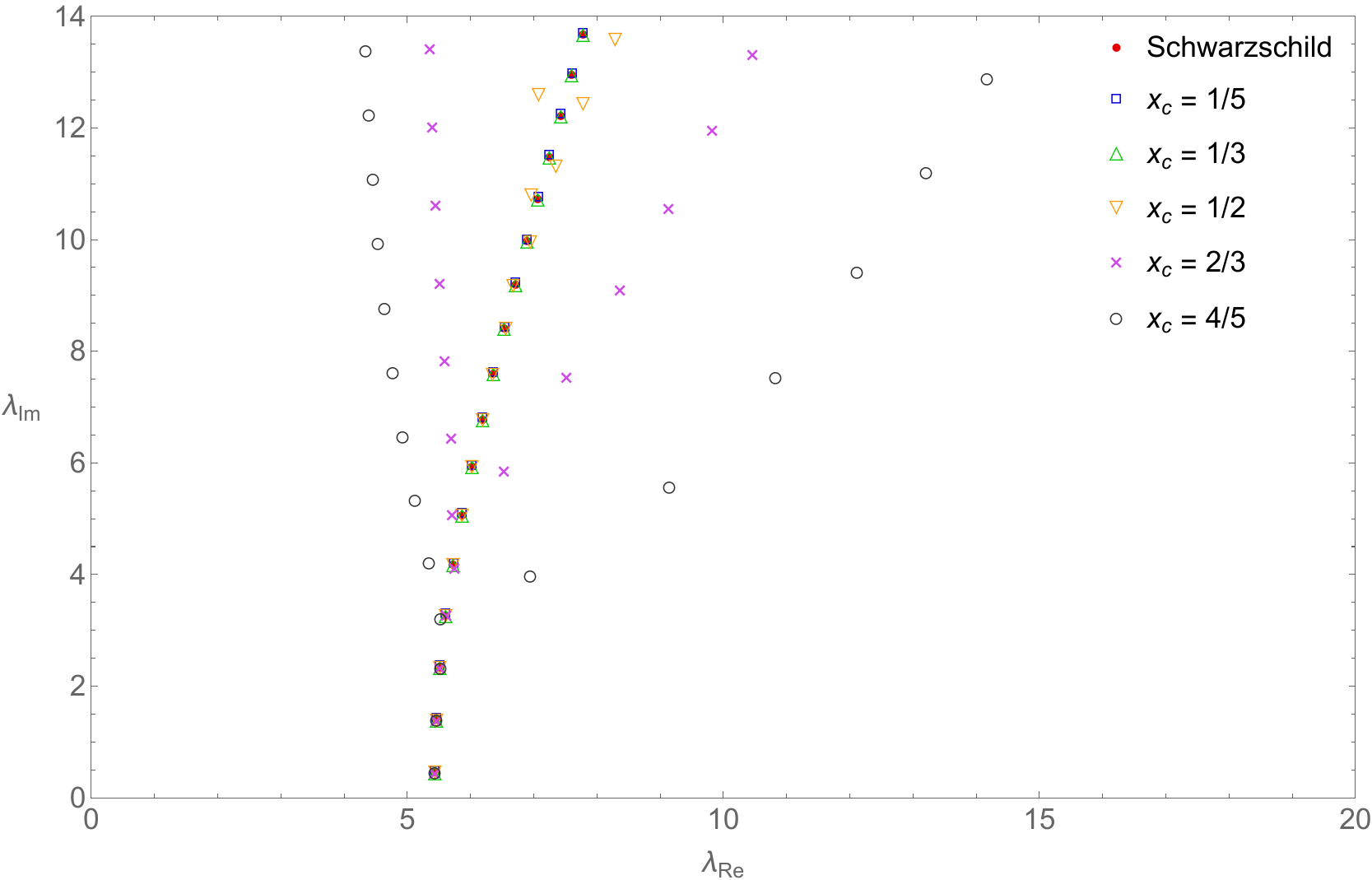}
\caption{The same as Fig.~\ref{fig01_RP_RWout} but evaluated at a larger frequency $\omega = 2.0$.}
\label{fig02_RP_RWout}
\end{figure}

The numerical results are presented in Tabs.~\ref{Tab.1}-\ref{Tab.3} and Figs.~\ref{fig01_RP_RWout} and~\ref{fig02_RP_RWout}.
In the tables, the Regge poles obtained using the improved matrix method are compared against those by the continued fraction method.
The latter utilizes an expansion of the wave function carried out around the truncation point and subsequently modifies the continued fraction method accordingly, as introduced in~\cite{agr-qnm-lq-matrix-06} for black hole QNMs.
Across these three tables, reasonable agreement between the two methods is observed.

The general tendency of the Regge poles is clearly shown by plotting the spectrum.  
By extending the calculation to include high-order Regge poles, the spectra are shown in Figs.~\ref{fig01_RP_RWout} and~\ref{fig02_RP_RWout} for the effective potential Eq.~\eqref{Veff_MRW3}.
For the sake of discussions elaborated in the next section, we consider two typical values of the frequency  $\omega=0.1$ and $\omega=2.0$.
Recalling the results in Refs~\cite{agr-qnm-instability-60} (c.f. Fig.~6 with $M=1/2$) and~\cite{agr-qnm-instability-61} (c.f. Fig.~3 with $r_0=1$), these two frequencies fall respectively into the relatively smaller frequency region where the greybody factor is found to be insensitive to the perturbation and the high-frequency region where sizable deviation from the unperturbed black hole is observed.
For minor frequencies, as shown in Fig.~\ref{fig01_RP_RWout}, the Regge pole spectra for the modified Regge-Wheeler potential remain very close to those of the unperturbed black hole, apart from some minor deviations observed in the higher-order modes.
From the perspective of Regge poles, one concludes that the stability of the greybody factor originates from those of the Regge pole spectrum.
However, for a more significant frequency, instability eventually occurs.
Specifically, a bifurcation merges in the spectrum, and the deviation from the unperturbed black hole initiates from high-order modes and propagates toward the low-lying one as the discontinuity moves away from the black hole. 
It is worth pointing out that this evolution of Regge poles is very similar to that of the QNM spectrum, observed recently in modified P\"oschl-Teller potential~\cite{agr-qnm-lq-matrix-12}.
Also, it is noted that the bifurcation in the Regge pole spectrum was first observed in~\cite{agr-qnm-Regge-13} for given frequency and metric perturbations.

Nonetheless, similar to QNMs, the Regge poles are not direct observables.
In this regard, it is meaningful to further evaluate the quantities directly associated with the observation, such as scattering amplitude and cross-section, elaborated earlier in the text.
Specifically, it was elaborated via numerical calculations in~\cite{agr-qnm-Regge-13} that the deformation of Regge poles might not significantly impact these observables.
This aspect will be explored in the following section. 

\section{Black hole scattering cross-section and spectral instability}\label{sec4}

\begin{figure}[htp]
\centering
\includegraphics[scale=0.5]{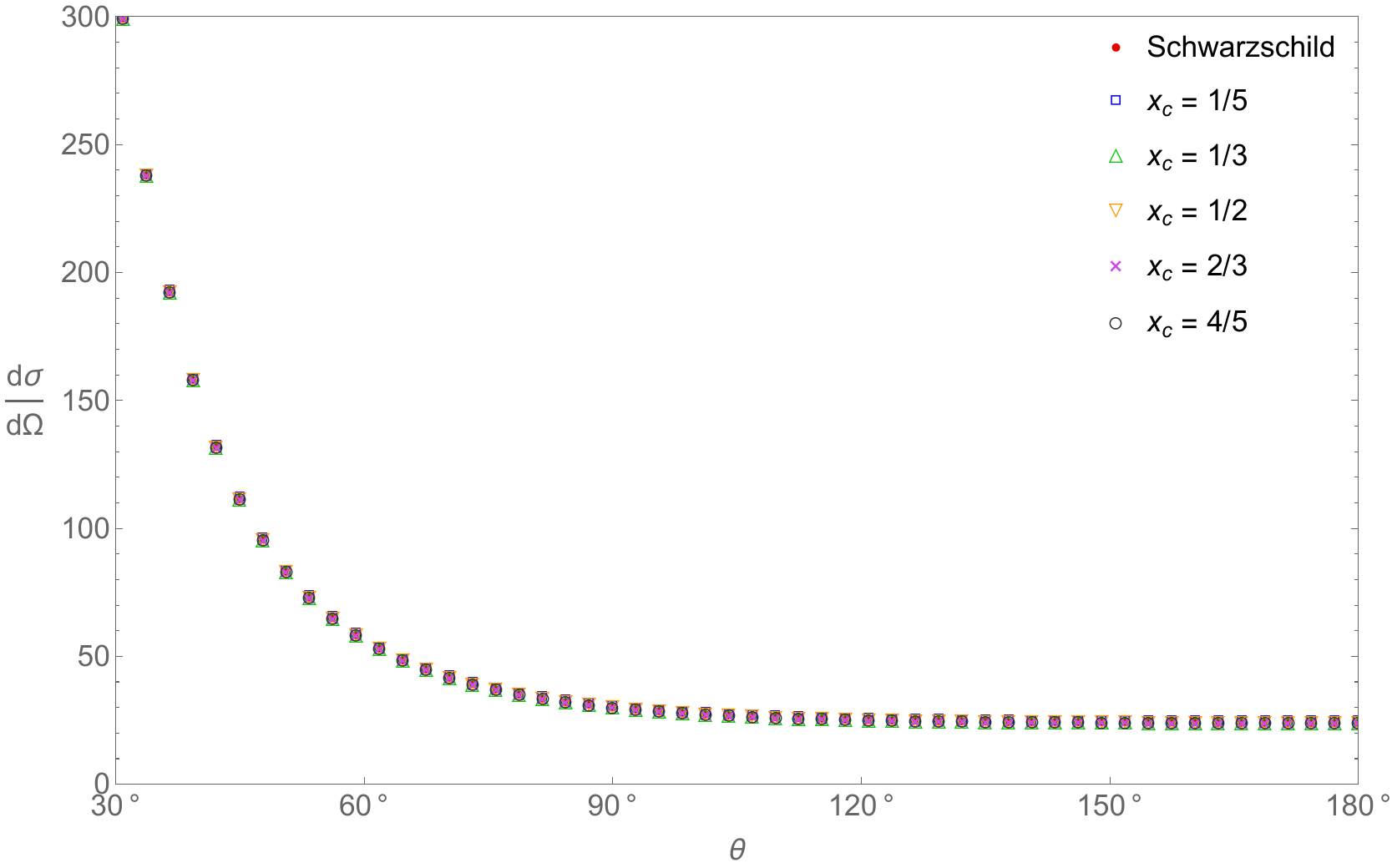}
\caption{The differential cross-sections as functions of the scattering angle obtained using the Regge poles for the modified Regge-Wheeler potential Eq.~\eqref{Veff_MRW3} with a minor step placed at different points $x_c= 1/5, 1/3$, $1/2$, $2/3$, and $4/5$.
The calculations are carried out for $\omega = 0.1$ by considering the first eight Regge poles.}
\label{fig03_RP_RWout}
\end{figure}

\begin{figure}[htp]
\centering
\includegraphics[scale=0.5]{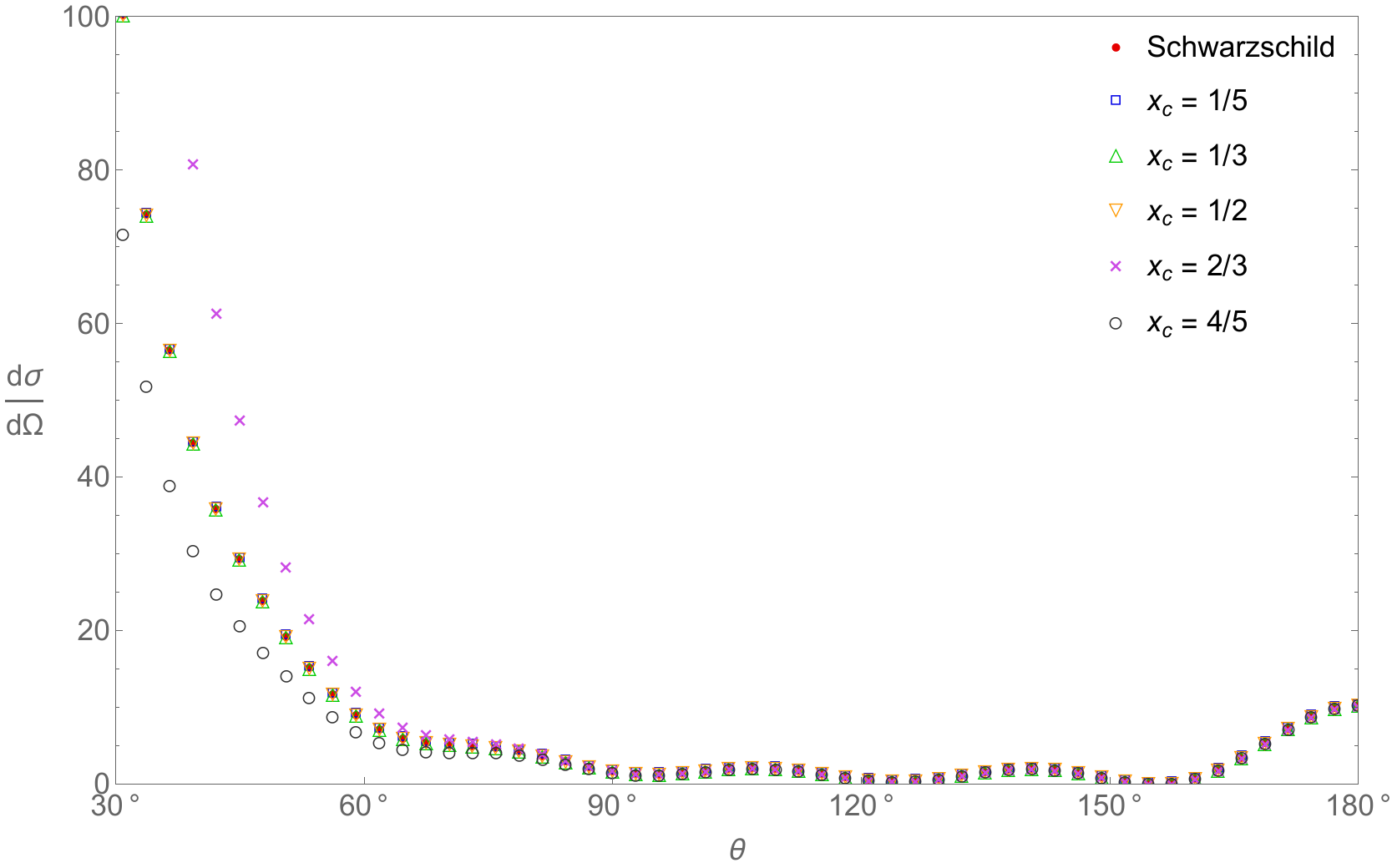}
\caption{The same as Fig.~\ref{fig03_RP_RWout} but evaluated for $\omega = 2.0$.}
\label{fig04_RP_RWout}
\end{figure}

This section elaborates on the observables, such as scattering amplitude and cross-section.
These quantities can be evaluated as summations of contributions from the residues of the Regge poles, according to Eqs.~\eqref{fRP} and~\eqref{sigmaRP}.
Again, we will primarily focus on the effective potential given by Eq.~\eqref{Veff_MRW3}, which corresponds to the metric perturbation due to a thin shell wrapped around the black hole.
The stability of the observables, such as the scattering amplitude and absorption cross-section, is analyzed in relation to the contributions from the Regge poles.
In particular, we analyze their dependence on the frequency and the location of the discontinuity.
The numerical results are shown in Figs.~\ref{fig03_RP_RWout} and~\ref{fig04_RP_RWout}.

At a moderate frequency, one observes in Fig.~\ref{fig03_RP_RWout} that the resultant differential cross-section is relatively insensitive to the location of the perturbation.
This is expected as the corresponding Regge poles shown in Fig.~\ref{fig01_RP_RWout} remain essentially unchanged.
Conversely, as shown in Fig.~\ref{fig04_RP_RWout}, at larger frequencies, the differential cross-section suffers a more significant deviation at small scattering angles.
One attributes the observed deviation to the more significant deformation in the high-order Regge poles shown in Fig.~\ref{fig02_RP_RWout}, which is triggered by ultraviolet metric perturbations.

\begin{figure}[htp]
\centering
\includegraphics[scale=0.5]{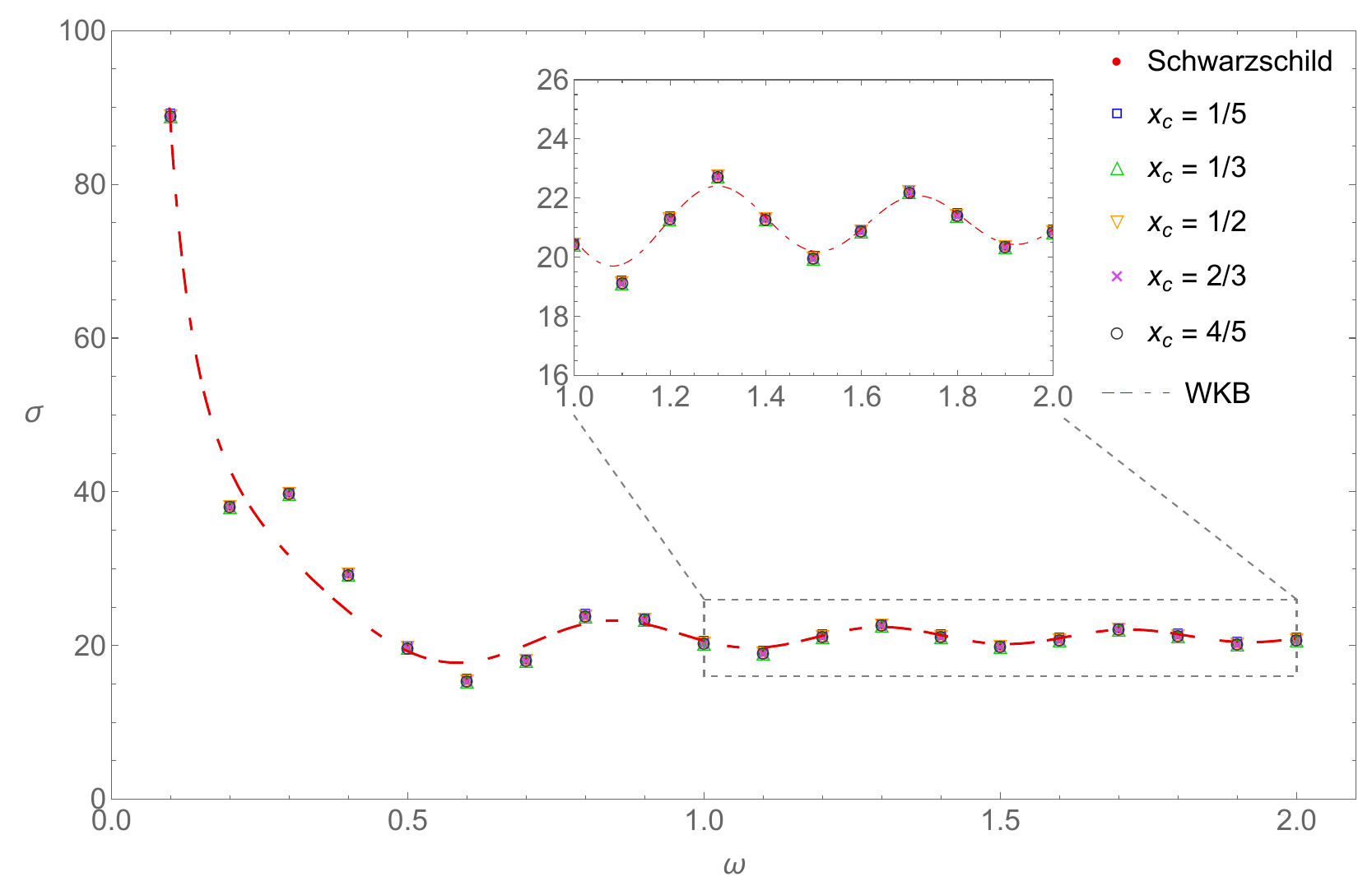}
\caption{The total cross-sections as functions of the frequency obtained using the Regge poles for the modified Regge-Wheeler potential Eq.~\eqref{Veff_MRW3} with a minor step placed at different points $x_c= 1/5, 1/3$, $1/2$, $2/3$, and $4/5$.
The calculations are carried out by considering the first eight Regge poles.
The results are compared against that obtained using the WKB approximation Eq.~\eqref{sigmaWKBRP} where only the first Regge pole of the unperturbed black hole is considered.
}
\label{fig05_RP_RWout}
\end{figure}

\begin{figure}[htp]
\centering
\includegraphics[scale=0.5]{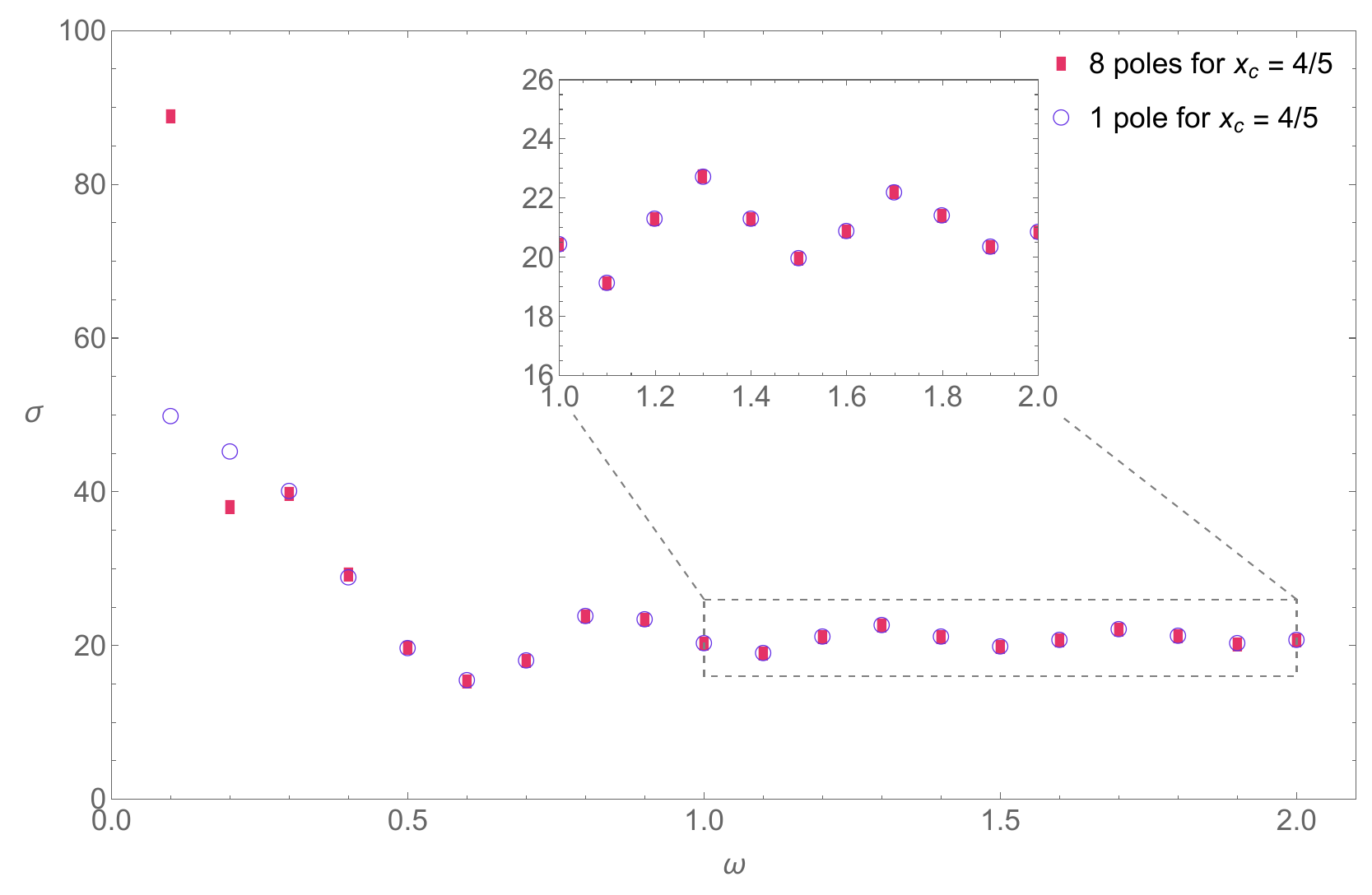}
\caption{The total cross-sections as functions of the frequency obtained using the Regge poles for the modified Regge-Wheeler potential Eq.~\eqref{Veff_MRW3} with a minor step placed at $x_c=4/5$.
The calculations are carried out by considering the first eight Regge poles and the first Regge pole.}
\label{fig06_RP_RWout}
\end{figure}

Before closing this section, we elaborate on an important connection between the Regge pole treatment and the WKB approximation.
Analogous to electromagnetic radiation, metric perturbations and, inclusively, the gravitational waves can be described by the WKB approximation at such a high energy limit~\cite{agr-qnm-Regge-09, agr-qnm-geometric-optics-06, agr-strong-lensing-correlator-15}.
It can be shown that, in this case, the absorption cross-section for a Schwarzschild black hole is primarily governed by the background and the low-lying Regge modes.
The former is nothing but the size of the black hole shadow obtained using geodesics.
The latter gives rise to additional oscillatory behavior as a function of the frequency, where the corrections from the high-order poles play an increasingly minor role.
Specifically, at the limit $\omega\to +\infty$, the primary contribution from the first Regge pole reads~\cite{agr-qnm-Regge-09, agr-qnm-Regge-10}
\bqn
\sigma^\mathrm{RP}_\mathrm{abs}(\omega) = -8\pi e^{-\pi}\left(\sigma_\mathrm{geo}+\frac{a_0 \pi}{\omega^2}\right) \mathrm{sinc}\left[2\pi\sqrt{27M^2\omega^2+a_0}\right] ,\lb{sigmaWKBRP}
\eqn
where
\bqn
a_0&=&\frac{2}{3}\left(\bar{s}^2-\frac{7}{72}\right),
\lb{sigmaBK}
\eqn
and
\bqn
\sigma_\mathrm{geo} &=& \pi r_\mathrm{c}^2 = 27\pi M^2 ,
\lb{sigmaBK}
\eqn
is the geometrical cross-section of the black hole where $r_\mathrm{c}=3\sqrt{3}M$ is the radius of the black hole shadow~\cite{agr-strong-lensing-shadow-review-03}.
It is noted that Eq.~\eqref{sigmaBK} corresponds to the background (of the contour) contribution.
The contribution to the total cross-section derived from the WKB approximation by taking account of various numbers of Regge poles and different locations of the mass shell is shown in~\ref{fig05_RP_RWout}, and~\ref{fig06_RP_RWout} where the result is presented as a function of frequency and compared against the unperturbed Schwarzschild case.
At high frequencies, the deviation from the Schwarzschild black hole is observed to be insignificant, consistent with the spirit of the WKB approximation.
The results are not sensitive to the location of the metric perturbation.
Also, the convergence in terms of Regge poles is surprisingly good, as the first Regge pole largely captures the essence of the entire correct.
At minor frequencies, Fig.~\ref{fig05_RP_RWout} indicates that deviation from the WKB approximation is observed, and Fig.~\ref{fig06_RP_RWout} implies that more Regge poles should be included.
Nonetheless, the overall stability remains robust, reinforcing the recent findings in Refs.~\cite{agr-qnm-instability-60, agr-qnm-instability-61}.

\section{Further discussions and concluding remarks}\label{sec5}

The challenge brought by the black hole spectral instability is that the substantial deviation in the QNM spectrum potentially leads to a sizable impact on the black hole spectroscopy, which utilizes the measured gravitational waves to extract the underlying parameters of the black holes.
In this regard, it was recently proposed~\cite{agr-qnm-instability-60, agr-qnm-instability-61} that the black hole greybody factors, or effectively their summation for different partial waves, the differential and total cross-sections, are relevant observables owing to their stability against small metric perturbations.
The findings imply that it might be more reliable to unambiguously extract the underlying black hole parameters from these observables.
In the mentioned studies, the instability of the QNMs was analyzed in contrast to the stability of the greybody factors, both of which were demonstrated by comparing against their counterparts of unperturbed Schwarzschild black holes. 
The greybody factors were primarily evaluated via direct numerical integration, and when necessary, high-order series expansions were employed to achieve the desired high precision.
By definition, the absorption cross-sections are related to the greybody factors by summing up all different partial waves, whose separation might not be straightforward from a practical viewpoint. 
Moreover, analogous to the fact that the time-domain waveforms can be viewed as a superposition of quasinormal oscillations, the absorption cross-sections are furnished by summing contributions from the Regge poles.

Primarily based on the above considerations, the present study aimed to explore the stability of the absorption cross-section via the analysis of Regge poles.
To this end, we generalize the matrix method to evaluate the Regge poles in black hole metrics with discontinuities. 
The obtained Regge pole spectrum is then used to calculate the scattering amplitude and cross-section. 
It is concluded that the stability of absorption cross-sections can be readily interpreted in terms of the Regge poles.
Specifically,
\begin{itemize}
\item Low-lying Regge poles are more stable compared to high-order ones. 
This is closely related to the fact that the phenomenon at high frequencies, where WKB approximation becomes feasible, is primarily governed by the background contribution, and contributions from the Regge poles provide mostly oscillatory corrections.
\item The observed instability is more vulnerable for high-order Regge poles, triggered by ultra-violet metric perturbations moving away from the black hole.
The bifurcation point, where instability emerges, is found to propagate from high-order poles to low-lying ones.
\item At large frequencies, some deviation from the unperturbed black hole metric is observed in the differential cross-section at small scattering angles, but the total cross-section remain stable and well-described by the WKB approximation.
\end{itemize}
Combining these findings leads to a better understanding of why greybody factors are more stable observables in terms of Regge poles.
On the one hand, the present study indicates that these modes become unstable only when the frequency is significant (for the setup employed in the present study, this occurs at $\omega\gtrsim 2$).
On the other hand, the validity of the eikonal limit implies that the corrections from the Regge poles, particularly the high-order ones, become minor at higher frequencies.
As a result, the overall effect from those unstable modes is suppressed, leading to more stable observables.

The present study is primarily focused on the first few low-lying Regge poles.
However, our findings encourage further analysis of the stability and eventual contribution of high-order Regge poles. 
More generally, the observational implications of this line of research primarily reside in their potential impact on gravitational wave detection.
Regarding QNMs, it has been speculated that the spectral instability might leave a detectable signature in the gravitational wave signals.
Specifically, these asymptotic QNMs align toward the real axis and might have a sizable collective effect.
The observable in question is the time-domain waveform where the frequency is integrated out.
The analysis regarding the greybody factor promotes a somewhat different perspective, where the observables are essentially frequency-domain scattering amplitudes, where the time dependence of the measurements is integrated out.
In this regard, it may be rather inviting to develop a unified approach where both types of singularity are brought into the same mathematical framework.
We plan to continue exploring these topics in future studies.

\section*{Acknowledgements}

We thank Michael D. Green and Stefan Randow for their insightful discussions.
We gratefully acknowledge the financial support from Brazilian agencies 
Funda\c{c}\~ao de Amparo \`a Pesquisa do Estado de S\~ao Paulo (FAPESP), 
Funda\c{c}\~ao de Amparo \`a Pesquisa do Estado do Rio de Janeiro (FAPERJ), 
Conselho Nacional de Desenvolvimento Cient\'{\i}fico e Tecnol\'ogico (CNPq), 
and Coordena\c{c}\~ao de Aperfei\c{c}oamento de Pessoal de N\'ivel Superior (CAPES).
This work is supported by the National Natural Science Foundation of China (NSFC).
GRL is supported by the China Scholarship Council.
A part of this work was developed under the project Institutos Nacionais de Ci\^{e}ncias e Tecnologia - F\'isica Nuclear e Aplica\c{c}\~{o}es (INCT/FNA) Proc. No. 464898/2014-5.
This research is also supported by the Center for Scientific Computing (NCC/GridUNESP) of S\~ao Paulo State University (UNESP).

\bibliographystyle{JHEP}
\bibliography{references_qian}

\end{document}